\documentclass[aps,prl,twocolumn,showpacs,preprintnumbers,amsmath,amssymb]{revtex4-1}

 \def\ep{{\epsilon}}

 \def\frac#1#2{{#1\over #2}}

 \def\s{\sqrt}

\def\be{\begin{equation}}
\def\ee{\end{equation}}
\def\ba{\begin{eqnarray}}
\def\ea{\end{eqnarray}}

 \def\de{\partial}

 \def\f {\frac}
 
 \def\ap{\alpha}

 \def\ddd{\cdot\cdot\cdot}
 \def\no{\nonumber \\}

 \def\la{\langle}
 \def\lb{\rangle}
 \def\ep{\epsilon}

\usepackage{color}
\usepackage{graphicx}
\usepackage{dcolumn}
\usepackage{bm}

\begin{document}

\title{Quantum Entanglement of Local Operators in Conformal Field Theories}
IPMU13-0244; YITP-13-132
\author{Masahiro Nozaki$^a$, Tokiro Numasawa$^a$ and Tadashi Takayanagi$^{a,b}$}

\affiliation{$^a$Yukawa Institute for Theoretical Physics,
Kyoto University, \\
Kitashirakawa Oiwakecho, Sakyo-ku, Kyoto 606-8502, Japan}

\affiliation{$^{b}$Kavli Institute for the Physics and Mathematics
 of the Universe,\\
University of Tokyo, Kashiwa, Chiba 277-8582, Japan}

\date{\today}

\begin{abstract}
We introduce a series of quantities which characterizes a given local operator in any conformal field theory from the viewpoint of quantum entanglement. It is defined by the increased amount of (Renyi) entanglement entropy at late time for an excited state defined by acting the local operator on the vacuum. We consider a conformal field theory on an infinite space and take the subsystem in the definition of the entanglement entropy to be its half. We calculate these quantities for a free massless scalar field theory in $2,4$ and $6$ dimensions. We find that these results are interpreted in terms of quantum entanglement of finite number of states, including EPR states. They agree with a heuristic picture of propagations of entangled particles.

\end{abstract}

\maketitle

{\bf{1. Introduction}}

Recently entanglement entropy has become a center of wide interest in a broad array of theoretical physics researches. It is defined as the von-Neumann entropy $S_A=-\mbox{Tr}[\rho_A\log \rho_A]$ of the reduced density matrix $\rho_A$
for a subsystem $A$. It has been used as a useful quantity which characterizes quantum properties of ground states in condensed matter physics (see e.g.\cite{capro,wen}).

 Moreover, it is intriguing to apply entanglement entropy to quantify excited states. For excited states in conformal field theories (CFTs), it was shown that entanglement entropy has an interesting property analogous to the first law of thermodynamics if the size of subsystem $A$ is much smaller than the excitation scale. This property was derived in \cite{thm} from the holographic entanglement entropy \cite{RT} and later a field theoretic derivation was given in \cite{BCHM}. Refer also to \cite{UAM} for an earlier related result.

Consider a CFT on a sphere times the time axis and pick up an excited state defined by acting a local operator $\mathcal{O}$ on the vacuum state $|0\lb$. Then the first law argues that the increased amount of entanglement entropy $\Delta S_A$  for this excited state, is essentially given by the conformal dimension of the operator $\mathcal{O}$ if the subsystem size (or equally the excitation energy) is very small.

On the other hand, it is natural to ask what will happen if we consider $\Delta S_A$ in the opposite limit i.e. the large size limit of subsystem $A$. One may expect that we get another basic quantity of an operator in CFTs which can be as fundamental as the conformal dimension. The main aim of this letter is to make a first step to answer this question. As we will see, this new quantity characterizes the quantum entanglement of an operator itself, together with its R$\acute{e}$nyi entropic versions.

There have been extensive studies on time evolutions of entanglement entropy in certain classes of largely excited states, called quantum quenches. One of them is called a global quench, which is triggered by changing parameters homogeneously \cite{cag} and is a special example of thermalization. Another class is called a local quench, which occurs by changing Hamiltonian locally \cite{cal,eis}.

{\bf{2. (R$\acute{e}$nyi) Entropy for locally Excited States}}

In this letter we focus on excited states which are defined by acting a local operators on the vacuum with a finite and positive conformal dimension in a given CFT. We consider a conformal field theory in the $d+1$ dimensional Euclidean space $R^{d+1}$, whose coordinates are
denoted by $(\tau, x^1,\ddd,x^{d})$. The density matrix $\rho$ for the total system is given by
$\rho=|\Psi\lb\la\Psi|$ and we choose the excited state $|\Psi\lb$ by acting an operator $\mathcal{O}$ as follows
\be
|\Psi\lb={\mathcal N} \cdot O(x^i)|0\lb,
\ee
where ${\mathcal{N}}$ is a normalization factor such that $\la\Psi|\Psi\lb=1$.
The constant $\mathcal{N}$ becomes finite after a proper regularization as we will explain later.
 Our state
$|\Psi\lb$ cannot be treated as a small perturbation from the vacuum state, though it describes an excited state much milder than that in local and global quantum quenches. Define also the ground state density matrix as $\rho_0=|0\lb\la 0|$.

To define the entanglement entropy, we choose the subsystem $A$ to be a half of the total space i.e. $x_1>0$. The reduced density matrix $\rho_A$ is defined by $\rho_A=\mbox{Tr}_B\rho$, tracing out the complement of $A$, called the subsystem $B$. The R$\acute{e}$nyi entanglement entropy $S^{(n)}_A$ is defined by
\be
S^{(n)}_A=\f{\log\mbox{Tr}[\rho_A^n]}{1-n}.
\ee
The limit $n\to 1$ coincides with the entanglement entropy $S_A$. The difference of $S^{(n)}_A$ between an excited state and the ground state is defined to be $\Delta S^{(n)}_A$.

We first calculate the entropies in the Euclidean formulation and finally perform an analytical continuation to see the dependence on the real time $t$. The time-evolution of density matrix is described by
\ba
\rho(t)&=&e^{-iHt}e^{-\ep H}\mathcal{O}(x_i)|0\lb\la 0|\mathcal{O}(x_i)e^{-\ep H}e^{iHt} \no
&=& \mathcal{O}(\tau_e)|0\lb\la 0|\mathcal{O}(\tau_l),
\ea
where we defined $\tau_e = -\epsilon - it, \ \ \tau_l = \epsilon - it$.
An infinitesimal parameter $\ep$ is an ultraviolet regularization.

In general, $\Delta S_A$ shows a non-trivial time-evolution.
Our analysis of explicit examples suggests that $\Delta S^{(n)}_A$ are monotonically increasing with the time $t$ for any local operator $\mathcal{O}$. Moreover, they finally approach to certain finite values $\Delta S^{(n)f}_A$ in the late time limit $t\to\infty$. These values  $\Delta S^{(n)f}_A$ depend on the choice of local operator $\mathcal{O}$ and are the quantities of our main interest.

\begin{figure}
  \centering
  \includegraphics[width=5cm]{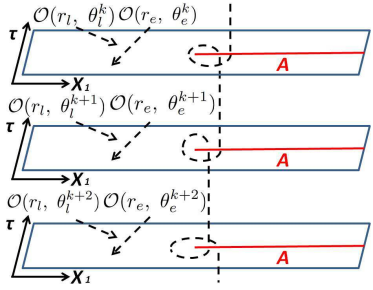}
  \caption{The $n$-sheeted geometry $\Sigma_n$ is constructed by gluing the upper cut along subsystem A on a sheet to the lower cut on the next sheet. }\label{rep}
\end{figure}

To calculate $S^{(n)}_A$, we employ the path-integral formalism by extending the replica method analysis in \cite{capro} for ground states. We can express
 Tr$\rho_A^n$ in terms of partition functions as Tr$\rho_A^n=Z_n/(Z_1)^n$. The partition on $R^{d+1}(=\Sigma_1)$, corresponding to $\la 0|O(\tau_l)O(\tau_e)|0\lb$, is written as
$Z_1$, while $Z_n$ is the partition function on $n$-sheeted space $\Sigma_n$ with
 $2n$ $\mathcal{O}$s inserted (see Fig.\ref{rep}). It is also useful to define the vacuum partition functions on $\Sigma_n$ and $R^{d+1}$ by $Z_{0n}$ and $Z_{01}$, respectively, so that we have  Tr$\rho_{0A}^n=Z_{0n}/(Z_{01})^n$ for the ground state.

\begin{figure}[hhh]
  \centering
  \includegraphics[width=5cm]{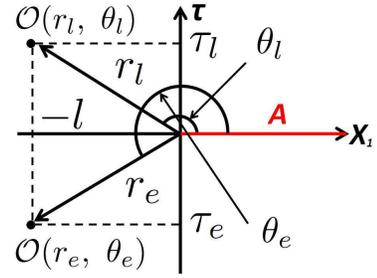}
  \caption{The Euclidean coordinate and operator insertions.}\label{pro2}
\end{figure}

In this way, we find that $\Delta S_A^{(n)}$ is rewritten as
\ba
&&\Delta S_A^{(n)} =\frac{1}{1-n}\left[\log{\frac{Z_n}{Z_{0n}}}-n\log{\frac{Z_1}{Z_{01}}}\right] \no
&&=\!\frac{1}{1-n}\Bigl[\log{\left \langle \mathcal{O}(r_l,\theta^n_l)\mathcal{O}(r_e, ,\theta^n_e)\!\cdots\! \mathcal{O}(r_l,\theta^1_l)\mathcal{O}(r_e,\theta^1_e) \! \right\rangle_{\Sigma_n}}\no
&&\ \ \ \ \ \ \ \ \ \ \ -n \log{\left\langle\mathcal{O}(r_l,\theta_l)\mathcal{O}(r_e,\theta_e) \right\rangle_{\Sigma_1}}\Bigr].
\ea
The term in the second line is given by a $2n$ points correlation function of $\mathcal{O}$ on $\Sigma_n$. The final term is a two point function of $\mathcal{O}$ on $R^{d+1}$. The values of
$r_{e,l}$ and $\theta^j_{e,l}$ are determined as follows. First we introduced the polar coordinate as $x_1+i\tau = re^{i\theta}$. The angular coordinate $\theta$ takes values $0\leq \theta <2n\pi$ on $\Sigma_n$ (see Fig.\ref{pro2}). We set $x_1=-l<0$ at each location of $\mathcal{O}$ and this measures the distance between the excited point and the boundary $\de A$ of the subsystem $A$. Since our calculations do not depend on locations in other directions $(x^2,\ddd,x^d)$, we omit their dependence. By defining $r_{e,l}\cdot e^{i\theta_{e,l}}\equiv -l+i\tau_{e,l}$, the $2n$ locations of the $\mathcal{O}$ insertions are given by $\theta^j_{e,l}\equiv \theta_{e,l}+2\pi(j-1),\ \ (j=1,2,\ddd,n)$.

{\bf 3. Results in Free Scalar Field Theories}

 To obtain analytical results of $\Delta S_A^{(n)}$, we focus on a free massless scalar field theory defined by the familiar action
$S=\int d^{d+1}x[(\de_{\tau}\phi)^2+(\de_{x_i}\phi)^2]$. We performed explicit calculations for various operators and replica numbers $n$ in 2, 4 and 6 dimension. We found that the results of late time values $\Delta S_A^{(n)f}$ do not depend on the dimension as long as the dimension is higher than two as we summarized in Table I. In two dimension, we will present results separately soon later. In this letter we will skip the details of the calculations because they are straightforward (but tedious) computations, employing the
 Green functions on $\Sigma_n$ in \cite{dow,sac}. Rather we give a brief summary of our results below. Refer also to the appendix A for Green functions and appendix B for some examples of entropy computations.


\begin{table}[hhh]\label{tr1}
\caption{$\Delta S^{(n) f}_A$ and $\Delta S^{f}_A\left(=\Delta S^{(1) f}_A\right)$ for free massless scalar field theories in dimensions higher than two ($d>1$).}
  \begin{tabular}{|c|c|c|c|c|c|} \hline
    & $n$ & $k=1$ & $k=2$ & $\cdots$ & $k=l$\\ \hline \hline
      & $2$ & $\log{2}$ & $\log{\frac{8}{3}}$ & $\cdots$ & $-\log{\left(\frac{1}{2^{2l}}\sum^l_{j=0}\left(_lC_j\right)^2\right)}$ \\ \cline{2-6}
   $\! \Delta\! S_A^{(n)\!f} $ & $3$ & $\log{2}$ & $\frac{1}{2}\log{\frac{32}{5}}$ & $\cdots$ & $ \frac{-1}{2}\log{\left(\frac{1}{2^{3l}}\sum^l_{j=0}\left(_lC_j\right)^3\right)}$\\ \cline{2-6}
      & $\vdots$ & $\vdots$ & $\vdots$ & $\vdots$ & $\vdots$\\ \cline{2-6}
& $m$ & $\log{2}$ & $\!\frac{1}{m-1}\!\log{\!\frac{2^{2m-1}}{2^{m-1}+1}}$ & $\cdots$ & $\!\frac{1}{1-m}\!\log{\!\left(\!\frac{1}{2^{m l}}\sum_{j=0}^{l}\!\left(_l C_j\right)^{m}\!\right)}$ \\ \hline
     $\! \Delta S^{f}_A$ & $1$ & $\log{2}$ & $\frac{3}{2}\log{2}$ & $\cdots$ & $\! l \log{2}\!-\!\frac{1}{2^l}\sum_{j=0}^{l}\!\! ~_lC_j\log{\!\! ~_lC_j}$ \\ \hline
  \end{tabular}
\end{table}

{\it Four and Six Dimensional Results}

First we describe the results in $4$ and $6$ dimensional case. As a series of local operators, we consider the primary operators
\be
\mathcal{O}=:\phi^k:\ \ \ (k=1,2,\ddd). \label{opo}
\ee
The time-evolutions of the R$\acute{e}$nyi entropies are all similar. In general, $\Delta S^{(n)}_A$ are vanishing in the region $t<l$. They start increasing at $t=l$ and keep to increase in the region $t>l$. Finally, they approach to certain constant values $\Delta S^{(n)f}_A$, in the late time limit $t\to \infty$.

\begin{figure}
  \centering
  \includegraphics[width=5cm]{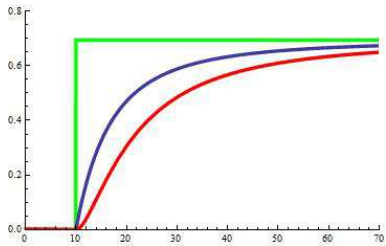}
  \caption{
The plots of $\Delta S^{(2)}_A$ as functions of $t$ in the limit $\ep=0$. We chose $l=10$. The red and blue curve correspond to the operator ${\mathcal{O}}=\phi$ ($k=1$) in 6 and 4 dimension, respectively. The green graph describes the entropy for the operator $\mathcal{O}=:e^{i \alpha \phi}:+:e^{-i \alpha \phi}:$ in 2 dimension.}\label{sample}
  \end{figure}

For example, the $n=2$ R$\acute{e}$nyi entropies for the operator $\mathcal{O}=\phi$ (i.e. $k=1$) in $4$ and $6$ dimension are given as follows when $t>l$, in the $\ep\to 0$ limit (see Fig.\ref{sample})
\ba
&&4d:\  \Delta S^{(2)}_A=\log \left(\f{2t^2}{t^2+l^2}\right), \label{renexp} \\
&&6d:\  \Delta S^{(2)}_A=\log \left(\f{8t^6}{l^6-6t^2l^4+9t^4l^2+4t^6}\right). \label{renexps}
\ea
Thus we find $\Delta S^{(2)f}_A=\log 2$. 

\begin{figure}[htbp]
 \begin{minipage}{0.4\hsize}
  \begin{center}
   \includegraphics[width=30mm]{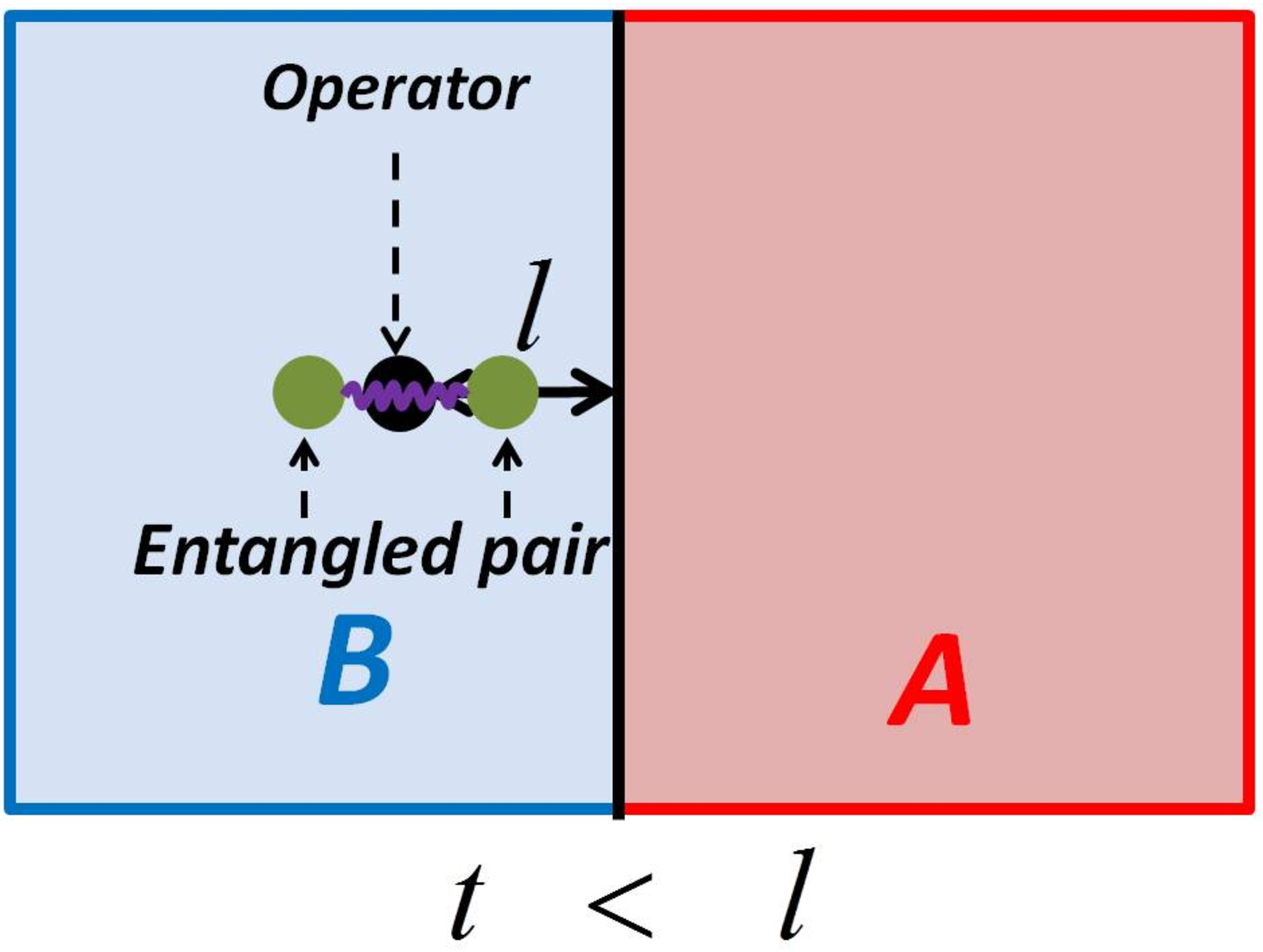}
  \end{center}
 \end{minipage}
  \begin{minipage}{0.4\hsize}
  \begin{center}
   \includegraphics[width=30mm]{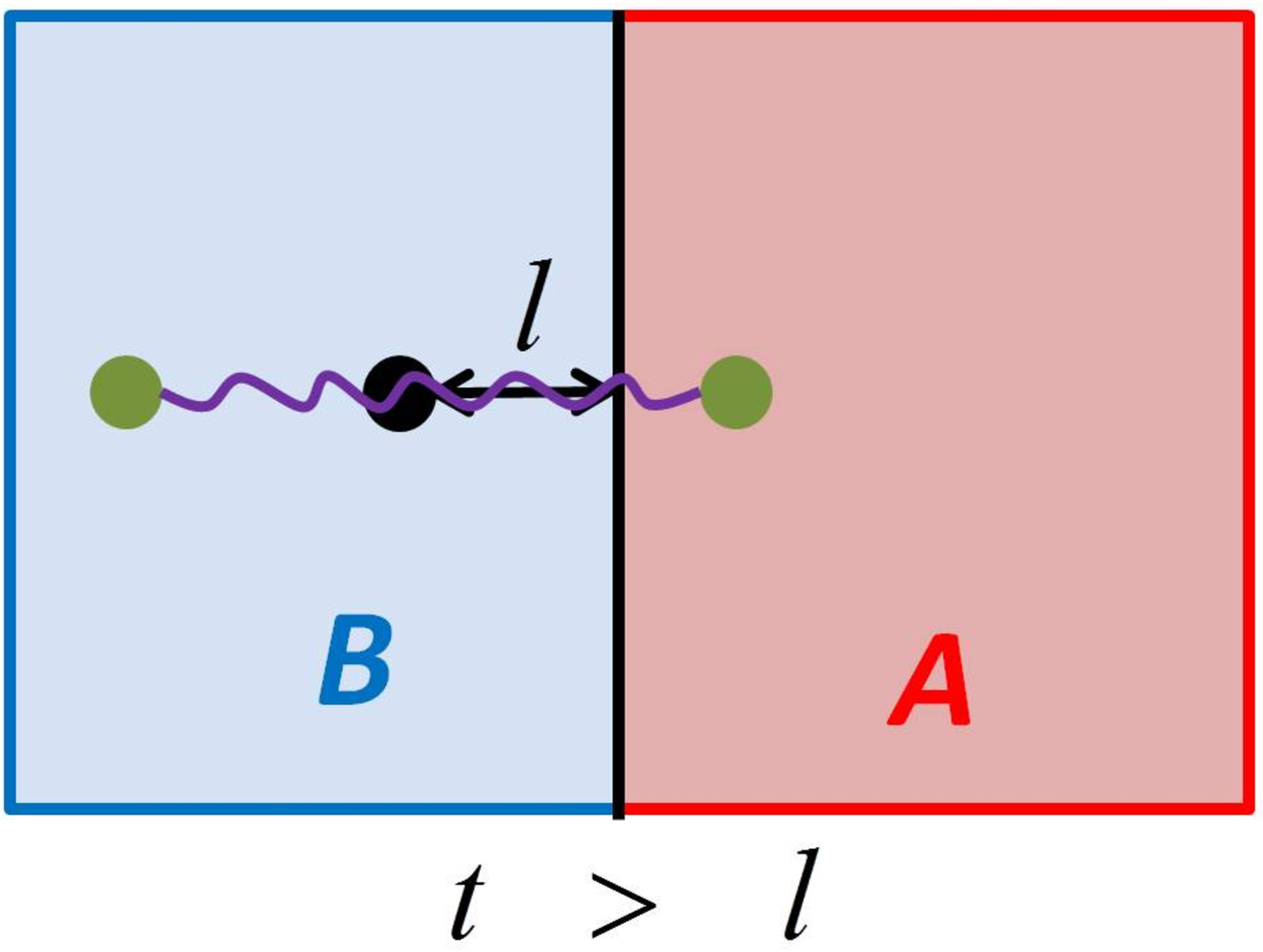}
  \end{center}
 \end{minipage}
 \caption{This shows a schematic explanation for the time evolution of $\Delta S^{(n)}_A$ in terms of entangled pairs.}\label{ph}
\end{figure}

We can interpret this behavior as follows. Firstly, an entangled pair of two quanta is produced at the point where the local operator is inserted as in Fig.\ref{ph}. And each of the two propagates in the opposite directions at the speed of light. When one of them reaches the boundary $\de A$ of the region A, the entanglement between two quanta starts to contribute to the (R$\acute{e}$nyi) entanglement entropies between A and B. Thus the minimum time for this event is $t=l$ and generally it takes more than $l$ for a pair moving in a generic direction. Finally, the entropies approach to certain constants as the two quanta remain to stay in $A$ and $B$, respectively, forever. This consideration also explains the monotonicity of entropy under time-evolutions.

 In the above, we did not take into account the conformal mass term $\propto \int_{\Sigma_n}R\phi^2$ of scalar field theory, where $R$ is the scalar curvature. Even though in general this affects results for excited states as noted in \cite{LM}, our results in the $\ep\to 0$ limit do not change by this effect.

{\it An Interpretation in terms of entangled pairs}

We calculated the late time values $\Delta S^{(n)f}_A$ for $\mathcal{O}=:\phi^k:$ with
various $(n,k)$ and we summarized them in Table I. Interestingly, we can find that $\Delta S^{(n)f}_A$ are equal to the values of R$\acute{e}$nyi entropies for $k+1$ dimensional Hilbert spaces under a simple rule. Indeed, let us define a reduced density matrix $\rho^{f}_A$ by the following $(k+1)\times (k+1)$ diagonal matrix
\be
\rho^f_A=2^{-k}({}_{k}C_{0}~,~{}_{k}C_{1}~,~\ddd~,~{}_{k}C_{k}), \label{frho}
\ee
where ${}_mC_{n}=\f{m!}{n!(m-n)!}$. Then we can confirm
\begin{equation}
\Delta S_A ^{(n)f} =\frac{1}{1-n}\log{\mbox{Tr}\left[ (\rho^f_A)^n\right]}.
\end{equation}

The R$\acute{e}$nyi entropy $\Delta S_A ^{(n)f}$ can be explicitly written as
\begin{equation}\label{aroe}
\Delta S^{(n)f}_A = \frac{1}{1-n}\log{\left(\frac{1}{2^{nk}}\sum_{j=0}^{k}~ (_{k}C_{j})^n\right)}.
\end{equation}
Taking the limit $n\to 1$ leads to the entanglement entropy
\begin{equation}\label{leen}
\Delta S_A= k \cdot \log{2}- \frac{1}{2^{k}}\sum _{j=0}^{k}~ _{k}C_{j}
\log{_{k}C_{j}}.
\end{equation}

Now we would like to provide an interpretation of the density matrices (\ref{frho}) in terms of
the entangled pairs. We decompose the scalar field as $\phi=\phi_L+\phi_R$, where $\phi_L$ and $\phi_R$ describe the modes which are moving toward the left ($x_1<0$) and right ($x_1>0$) direction. The key observation is that the late time entanglement entropy measures the entanglement between the left and right as $A$ is defined by $x_1>0$. We can expand our excited states as follows
\ba
|\Psi\lb=\mathcal{N}\cdot :\phi^n:|0\lb &=&\mathcal{N}\cdot\sum_{j=0}^k {}_kC_j\cdot (\phi_L)^j (\phi_R)^{k-j}|0\lb\no
&=&\!2^{-k/2}\!\sum_{j=0}^{k}\!\s{{}_kC_j}\!
\left|j\right\rangle_L\! \left|k-j \right\rangle_R,  \label{entstate}
\ea
where $|j\lb_{L,R}$ are normalized such that $\la i|j\lb_{L,R}=\delta_{i,j}$. Indeed, we can confirm that (\ref{entstate}) leads to the density matrix (\ref{frho}) after tracing out right-moving sector.

Especially, if we choose $k=1$, (\ref{entstate}) is equivalent to the maximally entangled
two $1/2$ spins (i.e. EPR state). Thus we find $\Delta S_A ^{(n)f}=\log 2$ for any $n$.
It might also be useful to notice that for $n=2$ we find the simple formula
\begin{equation} \label{2rel}
\Delta S^{(2)f}_A = \sum_{j=1}^{k}\log{\left[2j/(2j-1)\right]}.
\end{equation}

{\it Two dimensional Results}

Finally, we describe the results for the two dimensional free massless scalar.
As opposed to higher dimensions, the operators (\ref{opo}) cannot be regarded as local operators in our sense as their conformal dimensions are vanishing.

This motivates us to choose the following primary operators for any real values of $\ap$
\be
\mathcal{O}_1 =:e^{i \alpha \phi}:\ ,\ \ \ \ \mathcal{O}_2 =:e^{i \alpha \phi}:+:e^{-i \alpha \phi}:.
\ee

By explicit calculations, it is easy to show that  $\Delta S_A ^{(n)}$ and $\Delta S_A$ are always vanishing for the operator $\mathcal{O}_1$. On the other hand, if we consider the operator
$\mathcal{O}_2$, we obtain the following result
\ba
&& \Delta S_A ^{(n)}=\Delta S_A=0 \ \ \ (t<l),\no
&& \Delta S_A ^{(n)}=\Delta S_A= \log 2 \ \ \ (t>l).
\ea

Again we can explain these results in terms of the entangled pairs. In two dimension, we can  exactly decompose the scalar field into left and right-moving mode as $\phi=\phi_L(t+x_1)+\phi_R(t-x_1)$. Then it is obvious that the excited state $\mathcal{O}_1|0\lb=|e^{i\ap\phi_L}\lb_L|e^{i\ap\phi_R}\lb_R$ is a direct product state and should have the vanishing quantum entanglement. On the other hand,
\be
\mathcal{O}_2|0\lb=\left[|e^{i\ap\phi_L}\lb_L |e^{i\ap\phi_R}\lb_R+|e^{-i\ap\phi_L}\lb_L |e^{-i\ap\phi_R}\lb_R\right]/\s{2},
\ee
is the EPR state and has the entropy $\log 2$ for any $n$. Moreover, since in two dimension, the light-like motion of entangled pair is one dimensional and this is the reason why the entropy instantaneously jumps at $t=l$ as opposed to the results in higher dimensions (see Fig.\ref{sample}). It is curious to note that the results do not depend on the parameter $\ap$ or equally the conformal dimension.

The results for other local operators can be similarly understood in terms of the entangled pairs.  For example, consider operators of the form $\mathcal{O}=P(z)Q(\bar{z})$, where $P(z)$ and
$Q({\bar{z}})$ are arbitrary chiral and anti-chiral local operators such as
$\mathcal{O}_3=\de\phi\cdot \bar{\de}\phi$, where $\de$ and $\bar{\de}$ are the derivatives with respect to $z=x_1+i\tau$ and $\bar{z}=x_1-i\tau$. If we act these states on the vacuum $|0\lb$, it is obvious that they are all direct product states between left and right-moving sector. Therefore  $\Delta S_A ^{(n)}$ and $\Delta S_A$ are all vanishing when we take the cut off $\ep$ to be vanishing.

{\bf 4. Conclusions}

In this letter, we proposed a series of new quantities $\Delta S_A^{(n)f}$
which characterizes local operators in CFTs. In short, they measure the amount quantum entanglement of an operator or more intuitively quantum mechanical degrees of freedom included in an operator.

They are defined as the increased amount of
$n$-th R$\acute{e}$nyi entanglement entropy $\Delta S_A ^{(n)}$ at late time considering a
time-evolution of an excited state obtained by acting an operator on the vacuum. We chose the subsystem $A$ to be a half of the total space $R^{d}$. We conjectured that $\Delta S_A ^{(n)}$ are monotonically increasing functions of time.

We analyzed various explicit examples in free massless scalar field theories in $2, 4$ and $6$ dimension. They are enough to draw general conclusions for free massless scalar theories
in dimension higher than two, as summarized in Table I. We found that all of our results, even including two dimensional ones, can be understood in terms of quantum entanglement in finite dimensional Hilbert spaces like qubits in quantum information theory.

The behavior of $\Delta S_A ^{(n)}$ can be understood in terms of relativistic propagations of entangled pairs created by local operators. Indeed, the entropy starts increasing just when one of the entangled pair reaches the boundary $\de A$ of subsystem $A$. The time-evolution of entropy becomes step-functional in two dimension, while it gets a smooth function in higher dimension. This is because there are many directions to propagate and the arrival time at $\de A$ depends on directions in the latter.

Note that taking the subsystem $A$ to be infinitely large is important to obtain a non-zero constant entropy at late time. Our entangled pair interpretation suggests that the late time values $\Delta S_A^{(n)f}$ do not change even if we modify the shape of $A$ continuously. In this sense, they are topological quantities.

It is an interesting future problem to see how our results are changed in interacting CFTs, where our entangled quasi-particle interpretation might be modified. We may think of holographic computations similar to \cite{NNT}. It will also be intriguing to generalize our arguments to massive quantum field theories.

{\bf Acknowledgements} We thank J. Bhattacharya, S. He,
T. Nishioka, S. Ryu, N. Shiba and T. Ugajin for useful
discussions. TT is supported by JSPS Grant-in-Aid for Scientific
Research (B) No.25287058 and JSPS Grant-in-Aid for Challenging
Exploratory Research No.24654057. TT is also supported by World Premier
International Research Center Initiative (WPI Initiative) from the
Japan Ministry of Education, Culture, Sports, Science and Technology
(MEXT).

\appendix

\section{Appendix A: Propagators in n-sheeted space}

\begin{figure}[bbb]
  \centering
  \includegraphics[width=5cm]{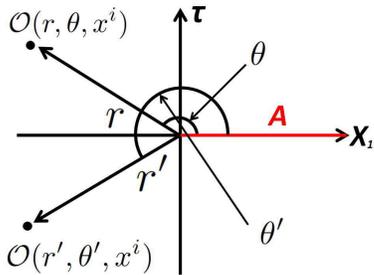}
  \caption{The subsystem A is defined by the region $x_1>0, \tau=0$. Operators are inserted into the points $(r, \theta, x^i ),~ (r', \theta ', x^i )$.}\label{pro1}
\end{figure}

We would like to compute R$\acute{e}$nyi entanglement entropies for locally excited states.
Then we have to construct propagators on an Euclidean space which has a conical singularity.
Below we consider the massless free scalar field theory on $n$-sheeted space $\Sigma_n$ in even dimensions. The shape of subsystem $A$ is chosen to be a half of the total space (see Fig.\ref{pro1}). We introduce the polar coordinate as $x_1+i\tau \!=\! re^{i\theta}$ $(0\le r<\infty, 0\leq \theta < 2 n\pi)$.

The Green function $G(r,r', \theta, \theta', {\bf x}, {\bf x'})$ is defined by
\begin{equation}
\begin{split}
&\mathcal{L}G(r,r', \theta, \theta', {\bf x}, {\bf x'})\!=\! \\
&\left(\partial_r^2+\frac{1}{r}\partial_r +\frac{1}{r^2}\partial_{\theta}^2+\partial_{\bf x}\right)G(r,r', \theta, \theta', {\bf x}, {\bf x'})\!=\! -\delta(x-x'),
\end{split}
\end{equation}
where ${\bf x}\!=\! (x^2, x^3, \cdots ,x^d)$.
Then we can expand the Green function by the eingenfunctions $u(r, \theta, {\bf x}) $,
\begin{equation}
\mathcal{L}u(r, \theta, {\bf x})\!=\!\lambda u(r, \theta, {\bf x}),
\end{equation}
where $u(r, \theta, {\bf x})\!=\!u(r, \theta+2n\pi, {\bf x})$.
As in \cite{capro}, the Green function can be expressed as
\begin{equation}\label{1}
\begin{split}
&G(r,r', \theta, \theta', {\bf x}, {\bf x'}) \\
&~~~~~~~\!=\!\frac{1}{2\pi n}\sum^{\infty}_{l\!=\!0}d_l \int ^{\infty}_0dk\int \frac{d^{d-1}k_{\perp}}{(2\pi)^{d-1}}\frac{k\cdot J_{\frac{l}{n}}(k r)J_{\frac{l}{n}}(k r')}{k^2+k_{\perp}^2} \\
&~~~~~~~~~~~~~~~~~~~~~~~~~~~~~~~~~~~\times e^{i{\bf k}_{\perp}\cdot \left({\bf x}'-{\bf x}\right)}\cos{\left(\frac{\theta-\theta'}{n}l\right)}.
\end{split}
\end{equation}
where $d_0 \!=\!1, d_{l>0}\!=\!2$ and $J_k(x)$ is the Bessel function of the first kind.
By using a Schwinger parameter, the denominator of (\ref{1}) can be expressed as
\begin{equation}
\frac{1}{k^2+k_{\perp}^2}\!=\!\int ^{\infty}_{0}ds e^{-(k^2+k_{\perp}^2)s}.
\end{equation}
We can integrate with respect to $k_{\perp}$,
\begin{equation}
\int \frac{d^{d-1}k_{\perp}}{(2\pi)^{d-1}} e^{-s(k_{\perp})^2+i{\bf k}_{\perp}\cdot \left({\bf x}'-{\bf x}\right)}\!=\!\frac{1}{(2\sqrt{s \pi})^{d-1}}e^{-\frac{1}{4s}\left({\bf x}'-{\bf x}\right)^2}.
\end{equation}
When we use the formula of Bessel function, the integral of $k$ from $0$ to $\infty$ in (\ref{1}) can be performed,
\begin{equation}
\int^{\infty}_{0}dk k e^{-s k^2}J_{\frac{l}{n}}(k r)J_{\frac{l}{n}}(k r')\!=\!\frac{1}{2s}e^{-\frac{r^2+r'^2}{4s}}I_{\frac{l}{n}}\left(\frac{rr'}{2s}\right).
\end{equation}
where $I_{\frac{l}{n}}(x)$ is the modified Bessel function of first kind.
And $I_{\frac{l}{n}}(x)$ is expressed as
\begin{equation}
I_{\frac{l}{n}}\left(\frac{rr'}{2s}\right)\!=\!\frac{1}{2i\pi}
\int^{\infty+i\pi}_{\infty-i\pi}dt~e^{\frac{rr'}{2s}\cosh{t}-\frac{l}{n}t}.
\end{equation}
After integrating with respect to $s$, $G(r,r', \theta, \theta', {\bf x}, {\bf x'})$ can be expressed as
\begin{equation}
\begin{split}
&G(r,r', \theta, \theta', {\bf x}, {\bf x'}) \!=\!\frac{\Gamma \left(\frac{d-1}{2}\right)}{4n\pi\left(2\pi^{\frac{1}{2}}\right)^{d-1}} \cdot\sum_{l\!=\!0}^{\infty}\frac{d_l}{2i\pi} \\
&\times \int^{\infty+i\pi}_{\infty-i\pi}dt\frac{4^{\frac{d-1}{2}}e^{-\frac{l}{n}t}\cdot \cos{\left(\frac{l\left(\theta-\theta'\right)}{n}\right)}}{\left\{\left(\left({\bf x}'-{\bf x}\right)^2+r^2+r'^2-2rr'\cosh{t} \right)\right\}^{\frac{d-1}{2}}}.
\end{split}
\end{equation}
When $d$ is odd, we can integrate with respect to $t$. When $d$ is $3$, $G(r,r', \theta, \theta', {\bf x}, {\bf x'})$ is given by
\ba
&& G(r,r', \theta, \theta', {\bf x}, {\bf x'}) \no
&&\!=\!\frac{1}{4n\pi^2 r r' (a-a^{-1})}\frac{a^{\frac{1}{n}}-a^{-\frac{1}{n}}}{a^{\frac{1}{n}}+a^{-\frac{1}{n}}-2\cos{\left( \frac{\theta-\theta'}{n}\right)} }.
\ea
where,
\begin{equation}
\begin{split}
\frac{a}{1+a^2}\!=\!\frac{r r'}{\left|{\bf x'}-{\bf x}\right|^2+r^2+r'^2},\\
\end{split}
\end{equation}
As in \cite{dow,sac}.
When $d$ is $5$, $G(r,r', \theta, \theta', {\bf x}, {\bf x'})$ is given by
\begin{equation}
\begin{split}
&G(r,r', \theta, \theta', {\bf x}, {\bf x'}) \!=\!\frac{f_n(a,\theta, \theta')+g_n(a,\theta, \theta')}{4\pi ^3 n^2 (r r')^2\left( a -a^{-1}\right)^2}, \\
\end{split}
\end{equation}
where $f_n, g_n$ are given by
\begin{equation}
\begin{split}
&f_n(a,\theta, \theta')\!=\!
\frac{2\left((a^{\frac{1}{n}}+a^{-\frac{1}{n}})\cos{\left(\frac{\theta-\theta'}{n}\right)}-2\right)}{\left(a^{\frac{1}{n}}+a^{-\frac{1}{n}}-2\cos{\left(\frac{\theta-\theta'}{n}\right)}\right)^2}, \\
&g_n(a,\theta , \theta' )\!=\!
\frac{n(a+a^{-1})(a^{\frac{1}{n}}-a^{-\frac{1}{n}})}{\left(a^{\frac{1}{n}}+a^{-\frac{1}{n}}-2\cos{ \left(\frac{\theta-\theta'}{n}\right)} \right) \left( a-a^{-1} \right) }.
\end{split}
\end{equation}

\section{Appendix B: Examples of Detailed Calculation of Entanglement Entropy}
We illustrate the detailed calculations of R$\acute{e}$nyi entanglement entropies in two, four and six dimensions by some examples.
\subsection{Two Dimension}
We calculate $\Delta S_A^{(2)}$ for $\left(:e^{i\alpha \phi(x)}:+ :e^{-i\alpha \phi(x)}:\right)\left|0\right\rangle$.
The explicit form of $\Delta S_{A}^{(2)}$ is given by
\begin{equation}
\begin{split}
&\Delta S^{(2)}_A \!=\!-\log\bigg{[}2^{-1-3 \alpha ^2} \\
&\times \left(8^{\alpha ^2}+(K_3)^{\frac{\alpha ^2}{2}} \left((-K_1+K_2)^{2 \alpha ^2}+(K_1+K_2)^{2 \alpha ^2}\right)\right)\bigg{]},\nonumber
\end{split}
\end{equation}
where $K_i$ are given by
\begin{equation}
\begin{split}
&K_1\!=\!\sqrt{l^2-t^2-\epsilon ^2+\sqrt{\left(l^2-t^2\right)^2+2 \left(l^2+t^2\right) \epsilon ^2+\epsilon ^4}},\\
&K_2\!=\!\sqrt{l^2-t^2+\epsilon ^2+\sqrt{\left(l^2-t^2\right)^2+2 \left(l^2+t^2\right) \epsilon ^2+\epsilon ^4}},\\
&K_3\!=\!\frac{1}{4 t^2 \epsilon ^2+\left(l^2-t^2+\epsilon ^2\right)^2}.\nonumber
\end{split}
\end{equation}
The time-evolution of $\Delta S_A^{(2)}$ is plotted in Fig.\ref{sample}.
The late-time constant value of $\Delta S_A^{(2)}$ is given by $\Delta S_A^{(2)f} \!=\!\log{2}$.

\subsection{Four Dimension}
We calculate $\Delta S_A^{(2)}$ for $\phi\left|0\right\rangle$ in four dimension.
Propagators which can contribute to $\Delta S_A^{(2)}$ are given by
\begin{equation}
\begin{split}
&\left\langle\phi(r_e,\theta_e)\phi(r_e,\theta_e+2\pi) \right\rangle_{\Sigma_2}\!=\!\frac{1}{64\pi^2r_e^2}, \\
&\left\langle\phi(r_l,\theta_l)\phi(r_l,\theta_l+2\pi) \right\rangle_{\Sigma_2}\!=\!\frac{1}{64\pi^2r_l^2}, \\
&\left\langle\phi(r_e,\theta_e)\phi(r_l,\theta_l) \right\rangle_{\Sigma_2}\!=\!\left\langle\phi(r_e,\theta_e+2\pi)\phi(r_l,\theta_l+2\pi) \right\rangle_{\Sigma_2} \\
&~~~~~~\!=\!\frac{1}{8\pi^2(r_e+r_l)\left(r_e+r_l-2\sqrt{r_e r_l}\cos{\left(\frac{\theta_e-\theta_l}{2}\right)}\right)}, \\
&\left\langle\phi(r_e,\theta_e+2\pi)\phi(r_l,\theta_l) \right\rangle_{\Sigma_2}\!=\!\left\langle\phi(r_e,\theta_e)\phi(r_l,\theta_l+2\pi) \right\rangle_{\Sigma_2} \\
&~~~~~~\!=\!\frac{1}{8\pi^2(r_e+r_l)\left(r_e+r_l+2\sqrt{r_e r_l}\cos{\left(\frac{\theta_e-\theta_l}{2}\right)}\right)}, \\
&\left\langle\phi(r_e,\theta_e)\phi(r_l,\theta_l) \right\rangle_{\Sigma_1}\!=\!\frac{1}{4 \pi^2\left\{r_e^2+r_l^2-2 r_e r_l \cos{\left(\theta_e-\theta_l\right)}\right\}}.\nonumber
\end{split}
\end{equation}
$\Delta S_A^{(2)}$ is given by the function of $(r_e, \theta_e, r_l, \theta_l)$.

After analytic continuation Euclidean signature to Lorentzian signature, the explicit form of $\Delta S_A^{(2)}$ is given by
\begin{equation} \label{d4n2s1}
\begin{split}
&\Delta S^{(2)}_A \!=\! -\log \bigg{[}\frac{l^2-t^2+\sqrt{(l^2-t^2+\epsilon^2)^2+4 t^2 \epsilon^2}}{l^2 -t^2 +\epsilon^2+\sqrt{(l^2-t^2+\epsilon^2)^2+4 t^2 \epsilon^2}} \\
&~~~~~~~~~~~~~~~~~~~~~~~~~~~~~~~+\frac{\epsilon^4}{16\left((l^2-t^2+\epsilon^2)^2+4 t^2 \epsilon^2\right)}\bigg{]}. \nonumber
\end{split}
\end{equation}

In the limit $\ep\to 0$, we find $\Delta S^{(2)}_A =0$ when $t<l$, while we obtain 
(\ref{renexp}) when $t\geq l$. The time-evolution of $\Delta S_A^{(2)}$ is plotted in Fig.\ref{sample}. Thus the late-time constant value of $\Delta S_A^{(2)}$ is given by
$\Delta S_A^{(2)f} \!=\!\log{2}$.

\subsection{Six Dimension}
We calculate $\Delta S_A^{(2)}$ for $\phi\left|0\right\rangle$ in six dimension.
Propagators which can contribute to $\Delta S_A^{(2)}$ are given by
\begin{equation}
\begin{split}
&\left\langle\phi(r_e,\theta_e)\phi(r_l,\theta_l) \right\rangle_{\Sigma_1}\!=\!\frac{1}{4\pi^3 \left(r_e^2+r_l^2-2r_e r_l \cos{(\theta_e-\theta_l)}\right)^2}, \\
&\left\langle\phi(r_e,\theta_e)\phi(r_e,\theta_e+2\pi) \right\rangle_{\Sigma_2} \!=\!\frac{3}{2^{10} \pi^3 r_e^4}, \\
&\left\langle\phi(r_e,\theta_e)\phi(r_e,\theta_e+2\pi) \right\rangle_{\Sigma_2} \!=\!\frac{3}{2^{10}\pi^3r_l^4}, \\
&\left\langle\phi(r_e,\theta_e)\phi(r_l,\theta_l+2\pi) \right\rangle_{\Sigma_2} \!=\!\left\langle\phi(r_e,\theta_e+2\pi)\phi(r_l,\theta_l) \right\rangle_{\Sigma_2} \\
&~~~~~~~~~~~~~\!=\!\frac{r_e+r_l+\sqrt{r_e r_l}\cos{\left(\frac{\theta_e-\theta_l}{2}\right)}}{8\pi^3 (r_e+r_l)^3 \left(r_e+r_l+2\sqrt{r_e r_l}\cos{\left( \frac{\theta_e-\theta_l}{2}\right)} \right)^2}, \\
&\left\langle\phi(r_e,\theta_e)\phi(r_l,\theta_l) \right\rangle_{\Sigma_2}\!=\!\left\langle\phi(r_e,\theta_e+2\pi)\phi(r_l,\theta_l+2\pi) \right\rangle_{\Sigma_2} \\
&~~~~~~~~~~~~~\!=\!\frac{r_e+r_l-\sqrt{r_e r_l}\cos{\left(\frac{\theta_e-\theta_l}{2}\right)}}{8\pi^3 (r_e+r_l)^3 \left(r_e+r_l-2\sqrt{r_e r_l}\cos{\left( \frac{\theta_e-\theta_l}{2}\right)} \right)^2}. \nonumber
\end{split}
\end{equation}
$\Delta S_A^{(2)}$ is given by the function of $(r_e, \theta_e, r_l, \theta_l)$.

After analytic continuation Euclidean signature to Lorentzian signature, the explicit form of $\Delta S_A^{(2)}$ is given by
\begin{equation}
\begin{split}
\Delta S^{(2)}_A \!=\!&\log\! \bigg{[} \!\left(\frac{9}{2^{20}\pi^6 A_4^4}\!+\!\frac{2A_2^3\!+\!9A_2^2A_3\!-\!12 A_2 A_3^2\!+\!4A_3^3}{64\pi^6 A_2^3 A_1^4}\right)^{-1} \\
&~~~~~~~~~~~~~~~~~~~~~~~~~~~~~~~~~~~~~~~~~~\cdot \left( \frac{1}{4\pi^3A_1^2}\right)^2\bigg{]},
\nonumber
\end{split}
\end{equation}
where $A_i$ are given by
\begin{equation}
\begin{split}
A_1&\!=\!4 \epsilon^2, \\
A_2&\!=\!2 (l^2-t^2+\epsilon^2)+2\sqrt{(l^2-t^2+\epsilon^2)^2+4t^2 \epsilon^2}, \\
A_3&\!=\! (l^2-t^2-\epsilon^2)+\sqrt{(l^2-t^2+\epsilon^2)^2+4t^2 \epsilon^2}, \\
A_4&\!=\!\sqrt{(l^2-t^2+\epsilon^2)^2+4t^2 \epsilon^2}. \nonumber
\end{split}
\end{equation}

In the limit $\ep\to 0$, we find $\Delta S^{(2)}_A =0$ when $t<l$, while we obtain
(\ref{renexps}) when $t\geq l$. The time-evolution of $\Delta S_A^{(2)}$ is plotted in Fig.\ref{sample}. Thus the late-time constant value of $\Delta S_A^{(2)}$ is given by
$\Delta S_A^{(2)f} \!=\!\log{2}$.

\end{document}